\documentclass[aps, prd, showpacs, floatfix, superscriptaddress, twocolumn, nofootinbib, reprint, longbibliography]{revtex4-2}
\usepackage{multirow, amsmath, hyperref, graphicx, booktabs}
\usepackage[T1]{fontenc}
\usepackage[utf8]{inputenc}
\usepackage{lmodern}
\usepackage{amsfonts}
\usepackage{amssymb}
\usepackage{url}
\usepackage{textcomp}
\usepackage{parskip}
\usepackage[dvipsnames]{xcolor}
\usepackage[autostyle]{csquotes}
\usepackage{orcidlink}
\usepackage[final]{microtype}
\hypersetup{
colorlinks=true,
citecolor=cyan,
linkcolor=magenta,
urlcolor=NavyBlue,
allbordercolors={0 0 0},
pdfborderstyle={/S/U/W 1}
}

\newcommand{\NIKHEF}{Nikhef, Science Park 105, 1098 XG Amsterdam, The Netherlands.}
\newcommand{\IWF}{Space Research Institute, Austrian Academy of Sciences, Schmiedlstrasse 6, 8042 Graz, Austria}
\newcommand{\CEICO}{CEICO, Institute of Physics of the Czech Academy of Sciences, Na Slovance 1999/2, 182 00, Prague 8, Czechia}
\begin{document}
\title{Mergers of hairy black holes: Constraining topological couplings from entropy}
\author{Kabir Chakravarti}
\email[]{chakravarti@fzu.cz}
\affiliation{\CEICO} 
\author{Amit Reza}
\email[]{amit.reza@oeaw.ac.at}
\affiliation{\NIKHEF} \affiliation{\IWF}
\author{Leonardo G. Trombetta}
\email[]{trombetta@fzu.cz}
\affiliation{\CEICO} 
\date{\today}
\begin{abstract}
Hairy black-holes are a unique prediction of certain theories that extend General Relativity (GR) with a scalar field. The presence of scalar hair is reflected non-trivially in the entropy of the black hole along with any topological coupling that may be present in the action. Demanding that a system of two merging black holes obeys the global second law of thermodynamics imposes a bound on this topological coupling coefficient. In this work we study how this bound is pushed from its GR value by the presence of scalar hair by considering estimates of binary black-hole merger parameters through inference studies of both mock and real gravitational-wave (GW) events. Although the scalar charge may produce a statistically significant deviation of the change in entropy over the GR prediction, we find no evidence of this happening in the data from real GW events taken from GWTC-1. We also find the entropy change to be susceptible to biases arising out of GW inferences which ends up being two orders of magnitude larger, therefore overwhelming any change, if at all, induced by the scalar hair.
\end{abstract}
\maketitle
\section{Introduction}
Among the most interesting predictions of General Relativity are the existence of black holes (BHs) and gravitational waves. The recent discovery of the latter \cite{LIGOScientific:2016aoc} has finally provided an opportunity to directly study black holes and their properties. One striking aspect of black holes is how they seemingly follow the laws of thermodynamics, made manifest by Bekenstein's identification of the horizon entropy with its area \cite{Bekenstein:1972tm}. The possibility of testing this feature is very compelling, in particular the second law associated to the increase of entropy across the merger process \cite{Isi:2020tac}. An equally enticing prospect of GW observations and the window they have opened of the strong-gravity regime is the possibility of also testing GR itself. A new laboratory of sorts. For this purpose it is necessary to have realistic expectations as to how deviations from GR can look like, and what kind of phenomenology they introduce. This requires the study of modified-gravity models and their black-hole solutions, as well as gravitational waves generated during the merger process. Among such models are of special interest those that predict that black holes are different from GR, for example by carrying a new charge, i.e. a violation of GR's no-hair theorem \cite{Bekenstein:1971hc}. 

A minimal extension of GR in this direction is achieved by scalar-tensor theories where hairy black-hole solutions have been found to exist, that is, black holes which carry a scalar charge, see Ref. \cite{Herdeiro:2015waa} for a review. These black holes are expected to exhibit phenomenological differences with respect to their GR counterparts. A particularly well-motivated extension of GR of this type is when a non-minimal coupling between the scalar field and the Gauss-Bonnet invariant is present, $f(\phi) \mathcal{G}$, also known as scalar-Gauss-Bonnet (sGB) gravity, where
\begin{equation}\label{GB-inv}
\mathcal{G} = R_{\mu\nu\rho\sigma} R^{\mu\nu\rho\sigma} - 4 R_{\mu\nu} R^{\mu\nu} + R^2 \,.
\end{equation}
This class of theories finds several motivations and has received a lot of attention in recent times. The associated phenomenology can be very rich and strongly depends on the form of the coupling function $f(\phi)$. On the one hand, an interesting situation is when there is no linear term, i.e. $f(\phi) = \lambda \phi^2 + \dots$, which enables the phenomenon of spontaneous scalarization, which is when both GR-like ($\phi=0$) and hairy solutions exist and a transition from one to the other can happen owing to a tachyonic instability induced by, depending on the sign of $\lambda$, either compactness \cite{Doneva:2017bvd,Silva:2017uqg} or spin \cite{Dima:2020yac}. On the other hand, when the coupling function starts with a linear term, \mbox{$f(\phi) = \alpha \phi + \dots$}, every black-hole carries secondary hair \cite{Sotiriou:2014pfa} while other horizon-less compact objects are not able to support any scalar charge \cite{Yagi:2015oca}. This particularlity makes it easier to constrain, as the ever-present scalar charge for BHs will inevitably generate dipolar radiation during the inspiral phase of a merger and introduce a dephasing in the GW signal which, so far, has not been observed and puts a bound of $\alpha < (1.2 \, \text{km})^2$ \cite{Lyu:2022gdr}. This bound already implies a small $\alpha \ll M^2$ regime, with $M$ the mass of the black hole. An example of a theory of this latter form is precisely the perturbative regime of Einstein-dilaton-Gauss-Bonnet (EdGB) gravity, where $f(\phi) = e^{\alpha \phi} - 1$, which can arise from the spontaneous breaking of a conformal symmetry \cite{Komargodski:2011vj}. Another special subcase is when the coupling is exactly linear $\alpha \phi \, \mathcal{G}$, i.e. no other terms are present at all, making the theory gain a shift-symmetry. When such symmetry is present there is a well known no-hair theorem for scalar-tensor theories \cite{Hui:2012qt}, which however finds its only exception precisely for this linear sGB coupling \cite{Sotiriou:2013qea}.

With black holes beyond GR it is also relevant to study their thermodynamics (see \cite{Sarkar:2019xfd} for a review). Thanks to work by Wald \emph{et. al.} \cite{Wald:1993nt,Iyer:1994ys} it is possible to define the horizon entropy for general theories of gravity by means of N\"other charges. This has been computed for the general sGB theories described above assuming a spherically symmetric, stationary black-hole horizon, giving \cite{Doneva:2017bvd}
\begin{equation} \label{eq:entropy-fsGB}
    S_\mathcal{H} = \frac{A_\mathcal{H}}{4} + 4\pi \gamma + \frac{4\pi}{\kappa} f(\phi_\mathcal{H}) \,,
\end{equation}
where $A_\mathcal{H}$ is the horizon area, $\phi_\mathcal{H}$ the value of the scalar field at the horizon and $\kappa = 1/16\pi$. The usual Bekenstein-Hawking formula is modified by a new term proportional to the coupling function evaluated at the horizon value of the scalar field. Importantly, there is also always a possible topological contribution proportional to $\gamma$, which is associated to a term in the action of the form $\kappa \gamma \, \mathcal{G}$. While such a term is non-dynamical in $D \leq 4$ dimensions, it still shows up in the entropy which makes it particularly relevant for black-hole thermodynamics, as the topology actually changes during black-hole formation through collapse, as well as during binary black-hole mergers. Indeed, a $\gamma > 0$ can increase the instability of de Sitter spacetime by favouring the nucleation of BHs \cite{Parikh:2009js}. On the other hand, violations of the second law can occur at the instant a black-hole horizon first forms during collapse, for $\gamma < 0$ \cite{Liko:2007vi, Sarkar:2010xp}. For these reasons it is important to bound the value of this type of topological coupling, as it has been explored in the case of GR extended with quadratic curvature invariants in \cite{Chakravarti:2022zeq}. 

In this paper we aim to find analogue bounds on $\gamma$ for the case of sGB theories containing a linear coupling term. For this purpose it important to distinguish between the approximately linear EdGB case and the exactly linear sGB one. The reason for this being the aforementioned shift-symmetry that emerges in the latter, which would be broken in a formula like Eq.~\eqref{eq:entropy-fsGB} unless it is appropriately fixed at the level of the action, including boundary terms \cite{Liska:2023fdz}. This will lead us to consider these two cases separately. In addition, realistic merger events involve spinning black holes, which requires to be accounted for by the proper generalization of the entropy formula for Kerr-like black holes. 

The paper is organized as follows: In Sec.~\ref{sec:entropy} we study how the Bekenstein-Hawking horizon entropy formula of GR is corrected when a linear coupling between the scalar field and the Gauss-Bonnet invariant is present in the action, accounting for the effect of rotation of a Kerr-like black hole. We make the distinction between the approximately and exactly shift-symmetric cases when appropriate. Then, in Sec.~\ref{sec:methodology} we outline our methodology to compute the entropy change before and after merger, along with a discussion on the caveats involved in our process. Finally, we discuss our results obtained on the changes bounds on the topological coefficient in Sec.~\ref{sec:reslts}.  

\section{Entropy in scalar-Gauss-Bonnet gravity}\label{sec:entropy}

We begin by considering a scalar-tensor theory where the scalar field is non-minimally coupled linearly to the Gauss-Bonnet invariant defined in Eq.~\eqref{GB-inv}, with action:
\begin{equation} \label{action-sGB}
    \mathcal{A} = \int d^4 \! x \, \sqrt{-g} \left[ \kappa \left(R + \gamma \, \mathcal{G} \right) - \frac{1}{2} (\partial \phi)^2 + \left(\alpha \phi + \dots \right) \mathcal{G} \right] \,,
\end{equation}
where $\kappa = 1/16\pi$ and we have also added a purely topological term with coupling constant $\gamma$ as discussed in the Introduction. In this form the theory can correspond to either EdGB or linear-sGB depending on the presence/absence of higher powers of $\phi$ in the ellipses above, and in fact in the small $\alpha \ll M^2$ regime we will consider, their dynamics are indistinguishable. Be it approximate or exact, the classical theory is then invariant under a constant shift of the scalar field $\phi \to \phi + C$. Indeed, this only generates a purely topological contribution $\alpha C \mathcal{G}$, which much like $\kappa \gamma \, \mathcal{G}$, gives just a boundary term. For this reason, neither $\gamma$ nor the value of $\phi$ itself (without derivatives) appear in the field equations, and therefore the dynamics remain insensitive to them. 

Due to the linear coupling between the scalar and the Gauss-Bonnet invariant in Eq.~\eqref{action-sGB}, in these theories the equation of motion for the scalar field always has a source term
\begin{eqnarray}\label{eq:phi}
    \square \phi + \alpha \mathcal{G} = 0 \, ,
\end{eqnarray}
meaning that the scalar field will have a non-trivial profile whenever $\mathcal{G} \neq 0$. As mentioned in the Introduction, in particular for a black-hole geometry a $\phi \simeq Q/r$ scalar hair is always generated with a secondary scalar charge $Q \sim \alpha/M$. This scalar profile will in turn induce changes in the geometry through backreaction with strength given by the dimensionless ratio $\alpha/M^2$. While exact solutions to Eq.~\eqref{eq:phi} are not known, one can still write approximate solutions which can be found numerically and/or perturbatively \cite{Sotiriou:2014pfa, Ayzenberg:2014aka}.

Boundary terms are however important for thermodynamical quantities such as the entropy. As it was shown by Wald, the entropy associated to a stationary black-hole horizon can be computed by:
\begin{eqnarray} \label{Wald-entropy}
    S_\mathcal{H} = \frac{2\pi}{\kappa} \int_{\mathcal{H}} \frac{\partial \mathcal{L}}{\partial R_{\mu\nu\rho\sigma}} \epsilon_{\mu\nu} \epsilon_{\rho\sigma} \,,
\end{eqnarray}
where $\epsilon_{\mu\nu}$ is the binormal tensor and the integration region $\mathcal{H}$ is a spatial cross-section of the horizon. For the theory in Eq.~\eqref{action-sGB} the above expression takes the explicit form \cite{Liska:2023fdz}:
\begin{eqnarray} \label{Wald-entropy-EdGB}
    S_\mathcal{H} &=& \frac{A_\mathcal{H}}{4} + 4\pi \gamma \\
    &&+ \frac{\alpha}{4\kappa} \int_{\mathcal{H}} \phi \left( \epsilon^{\mu\nu} R - 4 \epsilon^{\lambda \mu} R_\lambda{}^{\nu} + \epsilon_{\lambda\rho} R^{\nu\mu\lambda\rho} \right) \epsilon_{\mu\nu} d^2 \! A \,, \notag
\end{eqnarray}
where the first two terms come from the Einstein-Hilbert and topological terms in the action respectively, while the last piece is the contribution from the $\alpha \phi \, \mathcal{G}$ term. In order to explicitly compute the above expression for its later use in the case of realistic spinning black-holes, we must rely on the perturbative solutions of Ref.~\cite{Ayzenberg:2014aka} valid for small backreaction and slow rotation up to order $\alpha^2$ and $\chi^2$, where $\chi$ is the dimensionless spin parameter. As shown in Appendix~\ref{app:entropy}, this leads to
\begin{eqnarray}\label{eq:Dltn_final}
    S_\mathrm{EdGB} &=& \frac{A_\mathcal{H}}{4} + 4\pi \gamma + \frac{4 \pi \alpha^2}{\kappa M^2} \left( \frac{11}{6} - \chi^2 \frac{43}{240} \right) \, ,
\end{eqnarray}
where we are stressing that this result is only valid for the EdGB theory, i.e. when there is no exact shift-symmetry. The importance of boundary terms in the action becomes evident by the presence of the $\gamma$ in the entropy. The effects of the $\alpha$ coupling are two-fold: an explicit contribution that starts at order $\mathcal{O}(\alpha^2)$, and an implicit one inside $A_\mathcal{H}$ as the backreacting scalar shrinks the horizon size, also starting at $\mathcal{O}(\alpha^2)$ \cite{Ayzenberg:2014aka}. Both effects must be accounted for as they are of comparable size. Notice that our result, Eq.~\eqref{eq:Dltn_final}, is consistent with the spherically-symmetric expression given in Eq.~\eqref{eq:entropy-fsGB} for the case $f(\phi) \simeq \alpha \phi $, upon setting $\chi = 0$ and identifying $\phi_\mathcal{H} = 11\alpha/6M^2$ \cite{Sotiriou:2014pfa, Ayzenberg:2014aka}.

Turning to the linear-sGB case, as argued in Ref.~\cite{Liska:2023fdz} the entropy formula, Eq.~\eqref{Wald-entropy}, cannot be correct for an exactly shift-symmetric theory. This becomes immediately clear by performing a constant shift of the scalar field, which gives an extra $\Delta S_H = 4\pi \alpha C$ contribution, thus breaking the shift-symmetry at the level of the entropy. This traces back to the fact that the action in Eq.~\eqref{GB-inv} itself is only shift-symmetric up to a boundary term, making it not fully invariant although the field equations are. This suggests that a proper definition of a shift-symmetric action by the addition of a boundary term should lead to an entropy that respects the shift-symmetry everywhere. Following this prescription, the entropy for the linear-sGB theory instead reads 
\begin{eqnarray} \label{eq:Shft_final}
    S_\mathrm{Shift} &=& \frac{A_\mathcal{H}}{4} + 4\pi \gamma \, ,
\end{eqnarray}
which essentially differs from the EdGB entropy at Eq.~\eqref{eq:Dltn_final} in the removal of the explicit $\alpha$-contribution. Notice that the implicit $\alpha$-dependence is still present same as for EdGB, through the change in $A_\mathcal{H}$ with respect to GR as induced by the backreaction of the scalar \cite{Ayzenberg:2014aka}:
\begin{eqnarray}\label{eq:horizon-area}
    A_\mathcal{H} = A_{\mathcal{H},\mathrm{Kerr}} \left[ 1 - \frac{49}{40} \frac{\alpha^2}{\kappa M^4} \left( 1 + \frac{19}{98} \chi^2 \right) \right] \, ,
\end{eqnarray}
where $A_{\mathcal{H},\mathrm{Kerr}}$ is the horizon area for a Kerr black hole.

The computation of the entropy change across the merger follows from Eq.~\eqref{eq:horizon-area} along with Eqs.~\eqref{eq:Dltn_final} or \eqref{eq:Shft_final}, for EdGB or linear-sGB respectively. For this we are assuming that even though Wald's prescription given in Eq.~\eqref{Wald-entropy} requires stationarity, it still gives a good description of the entropy of the merging and final BHs in the asymptotic past/future. As we will describe in the following Section, GW data analysis then enables us to measure individual masses and spins before the merger from which we calculate entropy pre-merger, while appropriately found fitting functions infers the final mass and spin which calculates post-merger entropy. The entropy difference is thus
\begin{equation}\label{eq:entropy_diff}
    \Delta S = S^{(f)} - (S^{(1)}+S^{(2)}) \, ,
\end{equation}
where we are also assuming that the initial BHs can be treated independently when they are sufficiently separated, and therefore the total initial entropy can be expressed as the sum of the individual entropies. Furthermore, any loss of entropy through GW radiation is neglected. Importantly, the constant topological contribution proportional to $\gamma$ does not balance out since the initial and final configurations have different topology, i.e. two horizons merge into a single one. For this reason, in $\Delta S$ there is a contribution of the form $- 2\pi \gamma$ which offsets the change in entropy. In a pure gravity Effective Field Theory (EFT) valid up to the Planck scale $M_P$ it would reasonably be expected that this $\gamma$ term cannot ever lead to a violation of the second law for macroscopic BHs, i.e. $M \gg M_P$, which is when a thermodynamic description is sensible \cite{Chatterjee:2013daa}. However, the theory we are considering in Eq.~\eqref{action-sGB} has a significantly lower UV cutoff\footnote{The strong-coupling scale is at $\Lambda \sim (M_P/\alpha)^{1/3} \ll M_P$, while the theory in its vanilla form (without additional scalar operators) even requires a lower cutoff $\Lambda_{UV} \sim 1/\sqrt{\alpha}$ in order to avoid resolvable superluminal propagation \cite{Serra:2022pzl}.} and is therefore consistent with a much larger $\gamma$, making violations of the second law a priori possible within its regime of validity. Here we take the approach of demanding that $\Delta S \geq 0$ be satisfied during the merger process, allowing us to put an upper bound on the topological coupling $\gamma$.

\section{Method to compute entropy change}\label{sec:methodology}

The results of our computations from Sec.~\ref{sec:entropy}, especially Eqs. \eqref{eq:Dltn_final} and \eqref{eq:Shft_final} show that the entropy depends upon the masses and spins of the BHs is question as well as on the coupling coefficient $\alpha$. Estimating the entropy change in a BH merger thus requires along with $\alpha$ an estimation of the initial masses $M_1,M_2$ and spins $\chi_1,\chi_2$ as well as the mass and spin of the final configuration $M_f,\chi_f$. In what follows we will first briefly describe the waveform model in Sec.~\ref{ssec:waveforms} that we later use, then illustrate our methodology to obtain estimates on the component masses, spins and $\alpha$ (Sec.~\ref{ssec:bayes}). We will finally highlight our use of specific fitting functions in Sec.~\ref{ssec:fit} which will then be used to compute the entropy change. 

\subsection{Description of waveforms}\label{ssec:waveforms}

Our intended analysis will be inspiral-only and therefore is based on a parameterized post-Einsteinian (ppE) form, as was done in Ref.~\cite{Carson:2020cqb}. Accordingly our waveforms can be described by
\begin{eqnarray}\label{eq:ppe}
    \tilde{h} = A(u) \left(1 + \frac{f_a}{u^2}\right) \mathrm{exp}\left[ \Phi(u)\left(1 + \frac{f_p}{u^7}\right) \right] \, ,
\end{eqnarray}
where $u = (\pi\mathcal{M}f)^{1/3}$ is the inspiral PN-expansion parameter and $A(u),\Phi(u)$ are the GR contributions modelled after the $\texttt{IMRPhenomD}$ \cite{Husa:2015iqa} approximant. We use the conventional symbols $\mathcal{M}, \eta$ to denote the GW chirp-mass and the symmetric mass-ratio respectively. The presence of the non-minimal coupling has signatures in both the amplitude and the phase, although they come at different PN orders as seen in Eq.~\eqref{eq:ppe}. The quantities $f_a = -(5/192) \kappa$ and$f_p = (-5/7168) \kappa$ quantify the signatures, where $\kappa$ is given by
\begin{equation}\label{eq:kappa}
    \kappa = \frac{16\pi\alpha^2}{M^4} \frac{(m_1^2 s_2 - m_2^2 s_1)^2}{M^4\eta^{18/5}} \, ,
\end{equation}
with $s_i = 1 - \chi_i^2/4$. Once again as per convention, $m_i,\chi_i$ are used to denote component masses and spins of the binary whereas $M$ denotes the total mass.

To perform an inspiral-only analysis one must cutoff the waveform at some maximum frequency which normally is taken to be the frequency at the Last Stable Circular Orbit (LSCO). However the presence of the $\alpha$-dependent nonminimal coupling shifts the location and therefore the frequency of the LSCO. We then take the change in the LSCO radius from Ref.~\cite{Yunes:2016jcc}, add this to the GR LSCO radius and terminate at the corresponding $\alpha$-modulated frequency.

\subsection{Introduction to Bayesian inference}\label{ssec:bayes}

Let us briefly describe the general framework of Bayesian inference used to obtain the posterior distribution of GW signal parameters \cite{roulet2024inferring}. In GW data analysis, a specific detector's strain data $\boldsymbol{d}$ consists of an astrophysical GW signal $\boldsymbol{h}$ along with noise $\boldsymbol{n}$ and would be defined as  
\begin{equation}
\boldsymbol{d}^{(n)}(t) = \boldsymbol{h}^{(n)}(t|\theta) + \boldsymbol{n}^{(n)}  \, ,
\end{equation} 
where the superscript $n$ defines a specific detector from a network of detectors. In our work, we have assumed just the two LIGO detectors at Livingston and Hanford in the USA. $\boldsymbol{h}(t|\theta)$ defines the time-domain GW signal with a specific set of parameters $\theta$ (intrinsic and extrinisic). The intrinsic parameters are masses and spins for compact binary coalescences, and the extrinsic parameters are the coalescence phase, luminosity distance and the time of arrival at the geocentre. Concerning the geocentric reference frame, the strain measured at a detector of a GW source with polarization amplitudes $\boldsymbol{h}_{+}$, $\bf{h}_{\times}$ could be expressed as: 
\begin{equation}
\boldsymbol{h}^{(n)}(t|\theta) = F_{+}^{(n)}(i,j,k) \, \boldsymbol{h}_{+} +  F_{\times}^{(n)}(i,j,k) \, \boldsymbol{h}_{\times} \, ,
\end{equation}
where $F_{+, \times}$ denotes the antenna pattern functions (plus and cross) of the source locations $(i,j)$ and the polarization angle $k$ of the GW signal. $t$ denotes a specific time stamp which could be of $T = f_{s} \times \tau_{\text{obs}}$ seconds long. $f_{s}$ and $\tau_{\text{obs}}$ define the sampling frequency and observational time-window respectively. 

The estimation of the posterior probability density of the parameters $\boldsymbol{\theta}$ from the strain data of a specific detector could be obtained using Bayes' theorem as follows: 
\begin{equation}\label{eq:bayes}
p(\boldsymbol{\theta}|\boldsymbol{d}) \propto \mathcal{L}(\boldsymbol{d}|\boldsymbol{\theta}) \times \pi(\boldsymbol{\theta}) \, ,   
\end{equation}
where $\mathcal{L}(\boldsymbol{d}|\boldsymbol{\theta})$ and $\pi (\boldsymbol{\theta})$ define the likelihood and prior probability density function. The likelihood function computes the density function of the strain data corresponding to the unknown value of the parameters $\boldsymbol {\theta}$:
\begin{equation}\label{eq:log_L}
\ln (\mathcal{L} (\boldsymbol{d}|\boldsymbol{\theta})) = -\frac{1}{2} \, \langle \boldsymbol{d} -  \boldsymbol{h}(\theta) | \boldsymbol{d} -  \boldsymbol{h}(\theta) \rangle \, ,
\end{equation}
where the inner product $\langle . | . \rangle$ is defined as: 
\begin{equation}
\langle \boldsymbol{d}| \boldsymbol{h}  \rangle = 4 \, \mathrm{Re} \left[ \int{\frac{\tilde{d}(f) \, \tilde{h}^{*}(f) }{S_{n}(f)}} df \right] \, ,
\end{equation}
where, $S_{n}(f)$ denotes the one-sided power spectral density (PSD) of noise. $\tilde{h}(f)$ and $\tilde{d}(f)$ represent the frequency domain waveform and data respectively after performing the Fourier transform of time domain representation. The asterisk denotes the complex conjugate. 

In our studies we have considered a set of three simulated events (see Table \ref{Table:events}) as well as a set of three real events from GWTC-1 \cite{abbott2019gwtc} which are GW150914, GW170104, and GW170729. The strain data from the LIGO detectors for the real events have been taken from the GW Open Science Center (GWOSC) \cite{GWOSC}, while for the simulated cases BBH signals are injected into a two-detector network of LIGO-Hanford (LIGO-H) and LIGO-Livingston (LIGO-L) operating at design sensitivity to obtain the fake strain. For every case, we sample the log-likelihoods (Eq.~\eqref{eq:log_L}) using the $\texttt{bilby}$ library \cite{ashton2019bilby} and the $\texttt{dynesty}$ sampler. We then make use of a Bayesian inference (c.f. Eq.~\eqref{eq:bayes}) to infer the posterior probability distributions of masses and spin components along with $\alpha$.

\subsection{Fittings to obtain final mass and spin}\label{ssec:fit}

The estimation of final entropy $S_f$ requires the knowledge of post-merger BH configuration $M_f, \chi_f$. In principle, the observation of BH-ringdown can help us infer the final configuration as well. However, the amplitudes of the ringdown are exponentially damped and thus cannot be seen over the level of detector noise. To proceed one normally relies on numerical relativity (NR) simulations of BBH mergers. 
It turns out that NR simulations show consistent trends between the mass and spin $M_f,\chi_f$ of the final BH with the initial binary configuration $M_{1}, M_{2},\chi_{1}, \chi_{2}$, and can be fitted with appropriate fitting functions both in GR and in weakly coupled scalar-GB theories using the following relations presented in Ref.~\cite{Carson:2020cqb}:
\begin{equation}
M_{f} = M_{f, GR} + \zeta M_{f, \zeta} + \mathcal{O}(\zeta^{2}) \, ,
\end{equation}
\begin{equation}
\chi_{f} = \chi_{f, GR} + \zeta \chi_{f, \zeta} + \mathcal{O}(\zeta^{2}) \, ,
\end{equation}
Here expression for $M_{f, GR}$ and $\chi_{f, GR}$ have been taken from Ref. \cite{Husa:2015iqa}. $M_{f, \zeta}$ and $\chi_{f, \zeta}$ are the first order EdGB corrections and can be estimated as follows:
\begin{equation}
M_{f, \zeta} = M c_{0} \big(1 + c_{1} \chi_{f} + c_{2} \chi_{f}^{2} \big) + \mathcal{O}(\chi_{f}^{3}) \, ,
\end{equation}
\begin{equation}
\chi_{f, \zeta} = -d_{0} \eta \big(1 + d_{1} \chi_{f, GR} + d_{2} \chi_{f, GR}^{2} \big) + \mathcal{O}(\chi_{f, GR}^{3}) \, .
\end{equation}
Coefficients $c_{i}, d_{i}$ have been presented in Table 2 of Ref. \cite{Carson:2020cqb}. With this tool at our disposal, we can finally calculate both $A_\mathcal{H}^{(f)}(M_f,\chi_f)$ as well as $A_\mathcal{H}^{(1)}(M_1,\chi_1)$ and $A_\mathcal{H}^{(2)}(M_2,\chi_2)$, which can then be added with the appropriate $\alpha$ contributions to get the entropy once the inspiral estimates of masses and spins are known.    

\begin{figure}[ht!]
\centering
\includegraphics[width=1.0\columnwidth]{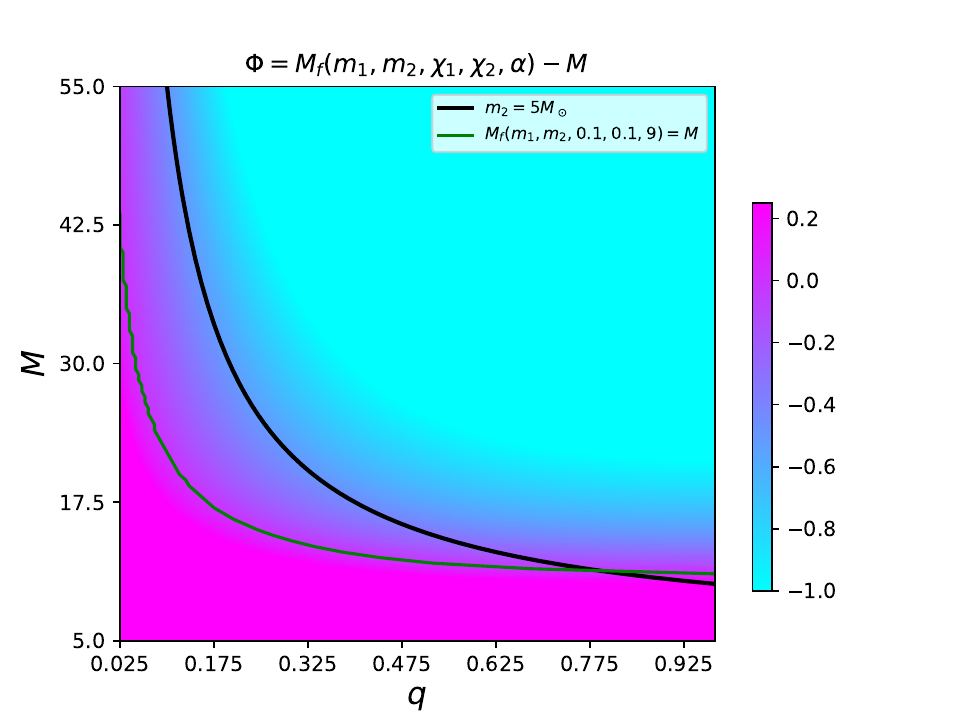}
\caption{Plot showing the values of $\Phi = M_f(m_{1}, m_{2}, \chi_{1}, \chi_{2}, \alpha) - M$ for binary configurations with spin values $\chi_{1} = \chi_{2} = 0.1$ and $\alpha = 9.0 \mathrm{[km]}^{2}$. Note that configurations with positive $\Phi$ are forbidden. The black line indicates $m_{2} = 5 M_{\odot}$.}
\label{fig:mlims}
\end{figure}

Before moving further it is important to pause and highlight a few caveats of our methods. To start, our entire framework makes ample use of perturbative stationary solutions \cite{Ayzenberg:2014aka} (see Appendix~\ref{app:entropy}) for both the metric $g_{\mu\nu}$ and the scalar field $\phi$. It is of importance for us to note that owing to the breakdown of these solutions, this framework cannot be applicable near the merger even though the corresponding expressions for the GR or $\phi$ coupled GW waveforms might still be true. Because of this, we have performed an inspiral-only analysis, where the BBH system can be well-approximated by two point-like sources and two non-interacting scalar fields. We remark that a higher-order perturbative solution could also make use of the full-length waveforms of \cite{Carson:2020cqb} including merger frequencies $f_\mathrm{RD},f_\mathrm{damp}$. The problem of $\phi$ dependent back-reaction is best evident in the fitting formulas for $M_f,\chi_f$ with $\alpha\neq0$. To be in the perturbative regime, one must have $\alpha \ll M^2$ or otherwise get unphysical results. This is depicted in Fig.~\ref{fig:mlims}. We plot the difference between the final mass $M_f$ and $M=M_1+M_2$, as a function of $M$ and $q = M_2/M_1 < 1$ with $\chi_1,\chi_2 = 0.1$ and $\alpha = 9.0 \mathrm{[km]}^2$. Interestingly Fig.~\ref{fig:mlims} shows the existence of forbidden regions which have $\Phi = M_f -M >0$ and therefore cannot be accessed with our choice of $\alpha/M^2$.     

\section{Results}\label{sec:reslts} 

As demonstrated in Sec.~\ref{sec:entropy} the respective expressions of Wald entropy for the dilaton and shift-symmetric coupling pick up their respective additional contributions dependent upon $\alpha$. Thus we see that in principle any bound on the parameter $\gamma$ should be affected by $\alpha\neq 0$. From the bounds on $\alpha$ obtained from BBH inspiral dynamics, we also know a priori that this change in the $\gamma$ explicitly caused by $\alpha$ will be small. Most of the budget of the entropy change before and after the merger will be dominated by the $\alpha=0$ Hawking-Bekenstein contribution, meaning that we can expect only a small deviation of the $\gamma$ bound from that found in \cite{Chakravarti:2022zeq}. Given this fact we want to answer two questions. First, we want to know if the $\alpha$ deviation to the $\gamma$ bound over the $\alpha=0$ baseline as predicted by GR is small enough to be insignificant. Provided that some non-zero values of $\alpha$ do cause a small but statistically significant shift to the $\gamma$ bound, it is interesting to compare this theoretical shift to the amount of biases. Biases are a well-known problem in GW templates which typically arise out of insufficiency in GW-template modeling \cite{Chakravarti:2018uyi}. We remind ourselves that any bias in GW templates will reflect themselves in biased estimates of the component masses and spins $(M_1,M_2,\chi_1,\chi_2)$ which would bias the estimate of the $\gamma$ bound {\it at the level of the GR contributions}. So as a second question we ask whether or not the biases dominate over the $\alpha$ changes.\\

\subsection{Quantifying the effect of $\alpha$}\label{ssec:alpha}

\begin{figure}[ht!]
    \centering
    \includegraphics[width=1.2\columnwidth]{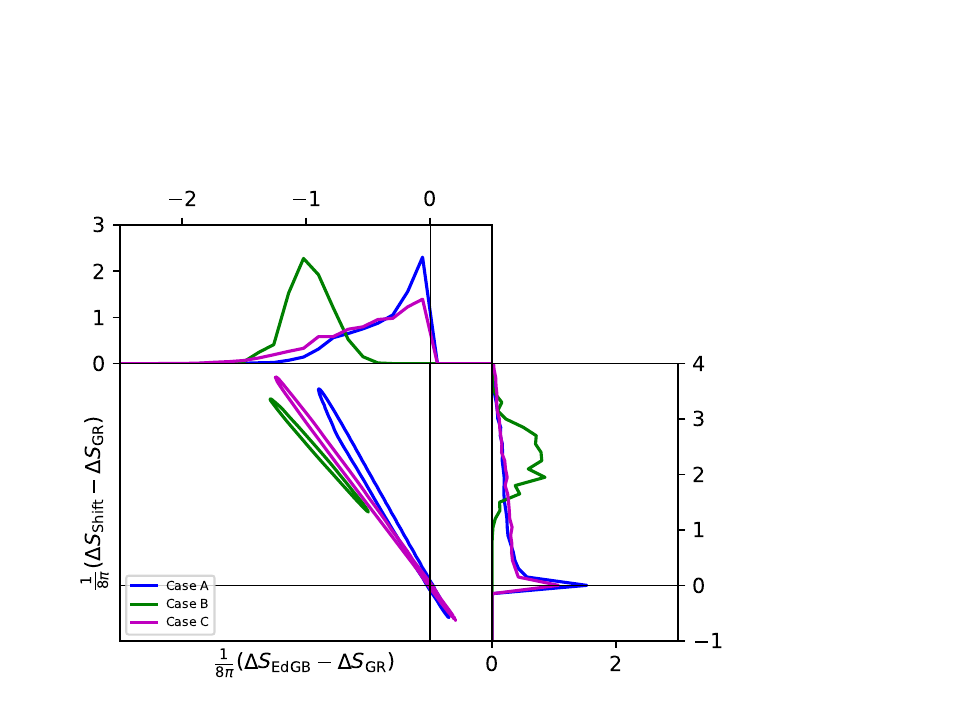}
    \caption{Plot showing the histogram of values of $\delta(\Delta S)$ for the dilaton and the shift symmetric cases, for the simulated GW events of Table.~\ref{Table:events}. Note that we cannot claim a deviation statistically significant from the GR baseline if the $90\%$ CLs of $\delta(\Delta S)$ touch the $\delta(\Delta S) = 0$ lines.}
    \label{fig:schange}
\end{figure}

\begin{figure}[ht!]
\centering
\includegraphics[width=1.2\columnwidth]{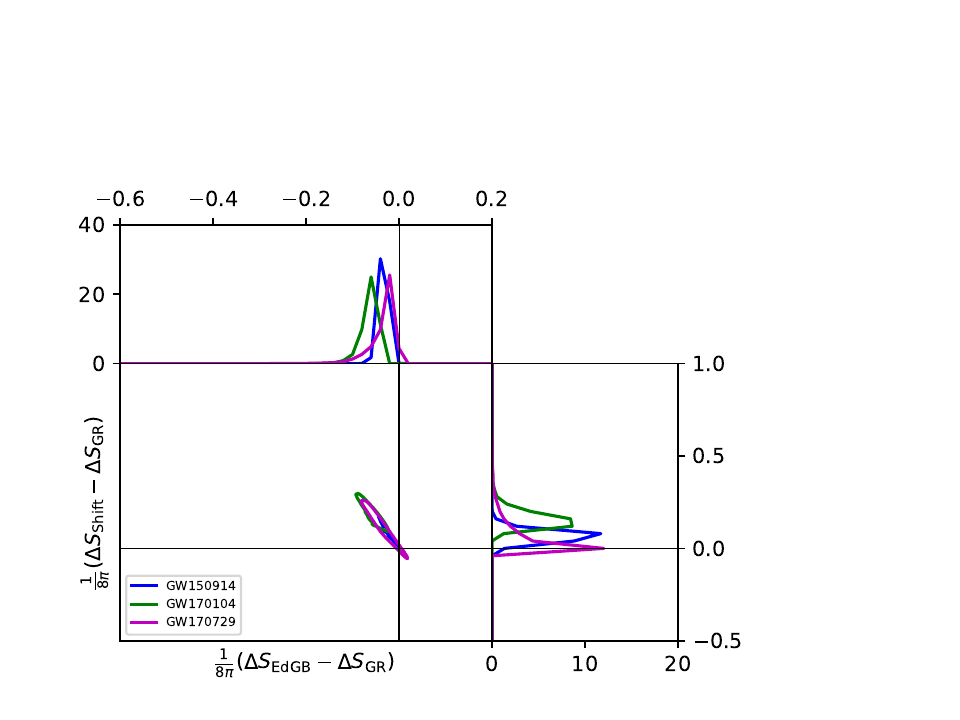}
\caption{Same as Fig.~\ref{fig:schange} except we use the real events GW150914, GW1701104, and GW170729 \cite{abbott2019gwtc, GWOSC}.}
\label{fig:schange-real}
\end{figure}

In order to understand the consequences of the $\alpha$-dependent coupling we devise two separate procedures involving separate instances of the GW events. For the first procedure, we consider both mock and real events. The injections for the mock events involve each of Case A, Case B and Case C as given in Table \ref{Table:events}, along with an $\sqrt{\alpha}=3\mathrm{[km]}$. We choose Case A as a GW150914-like \cite{abbott2016properties} event with binary BH masses of $36.0 M_\odot$ and $29.0 M_\odot$ with a signal SNR $\rho \sim 15$. Then Case B and Case C are arbitrarily chosen events with stellar mass BHs and realistic SNRs. Thus Case A, Case B and Case C along with the GW150914, GW170104 and GW170729 make up all the cases considered in this procedure. Making use of the waveforms of \cite{Carson:2020cqb}, we use the Bayesian method (c.f  Sec.~\ref{ssec:bayes}) to infer $M_i,\chi_i$ and $\alpha$ for all the cases considered. Then we computed the difference in entropy change (before and after the merger) between both the dilaton and shift symmetric expressions and the Hawking-Bekenstein one. In other words we have computed 
\begin{align}
    \delta(\Delta S_\mathrm{EdGB}) &= \Delta S_\mathrm{EdGB} - \Delta S_\mathrm{GR}  \\
    \delta(\Delta S_\mathrm{Shift}) &= \Delta S_\mathrm{Shift} - \Delta S_\mathrm{GR} 
\end{align} 
where $\Delta S$ is the respective `EdGB', `Shift' or GR merger entropy change from Eq.~\eqref{eq:entropy_diff}. We remind ourselves that in the non-GR definitions of $\Delta S$ here, we do not factor in the $\gamma$ effect from the boundary (c.f. Eq.~\eqref{eq:Dltn_final} and Eq.~\eqref{eq:Shft_final}). Fig.~\ref{fig:schange} shows the simultaneous plots of $\delta(\Delta S_\mathrm{EdGB})$ and $\delta(\Delta S_\mathrm{Shift})$ for the set of simulated GW events while Fig.~\ref{fig:schange-real} shows the same for the set of real events we considered.

\begin{table}
\begin{center}
\begin{tabular}{||c| p{2.2cm}| c| c| c| c| c||} 
 \hline
 \# & Label & $m_1$ & $m_2$ & $\chi_1$ & $\chi_2$ & $\rho$(SNR) \\ [0.5ex] 
 \hline\hline
 1& Case A & 36.0 & 29.0 & 0.4 & 0.3 & 14.89 \\ 
 \hline
 2& Case B & 58.0 & 15.0 & 0.02 & 0.06 & 27.86 \\
 \hline
 3& Case C & 50.0 & 20.0 & 0.4 & 0.3 & 26.07 \\
 \hline
\end{tabular}
\end{center}
\caption{Table showing the configurations for our choices of simulated GW events. For each of these events we assume $\alpha = 9.0 \mathrm{[km]}^2$. Units of $m_{1}, m_{2}$ are in $M_{\odot}$, while the SNRs are the network SNRs over LIGO-HL network.}
\label{Table:events}
\end{table} 
The results from both our simulated cases and real events considered lead to several interesting conclusions. To begin with, we note that the 2-d distributions (Figs.~\ref{fig:schange} and \ref{fig:schange-real}) that the $95\%$ CLs of the $\delta(\Delta S)$ lies either completely or overwhelmingly in the  $\delta(\Delta S_\mathrm{EdGB})<0 $ and $\delta(\Delta S_\mathrm{Shift})>0$ quadrant. Statistically it just means that the data from the cases considered show a greater odds-ratio favouring $\delta(S_\mathrm{EdGB})<0 $ and $\delta(\Delta S_\mathrm{Shift})>0$.  However, it should be noted that a statistically significant deviation above the GR baseline model for either of the theories can only be proven for a particular event if the $90\%$ CL of that event do not touch the corresponding $\delta(\Delta S)=0 $ baseline. In this regard, the simulated event labeled Case B stand out as the one able to predict statistically significant deviation at an injection of $\alpha = 9.0 \mathrm{[km]}^2$, whereas Case A and Case C do not. This also is expected, as we see from Eq.~\eqref{eq:Dltn_final} and Eq.~\eqref{eq:Shft_final} that the expression of the Wald entropy depends upon the dimensionless factor $\alpha^2/M^4$. Consequently these events which had a proportionately bigger contribution owing to the lighter BHs was able to give us a statistically significant deviation. In other words, we see the presence of a dynamical coupling to a scalar field through the $\alpha$ affects the entropy computation and in principle can show up even in the histograms computed for the entropy. However, it should also be remembered that the magnitude of the shift in entropy change is only $\sim\mathcal{O}(1)$, which is only a very tiny fraction of the $\sim\mathcal{O}(100)$ entropy change caused by GR. This is not surprising given that the perturbative regime we chose means that $\alpha$ effects cannot dominate over the GR ones.

\begin{figure}[ht!]
    \centering
    \includegraphics[width=1.2\columnwidth]{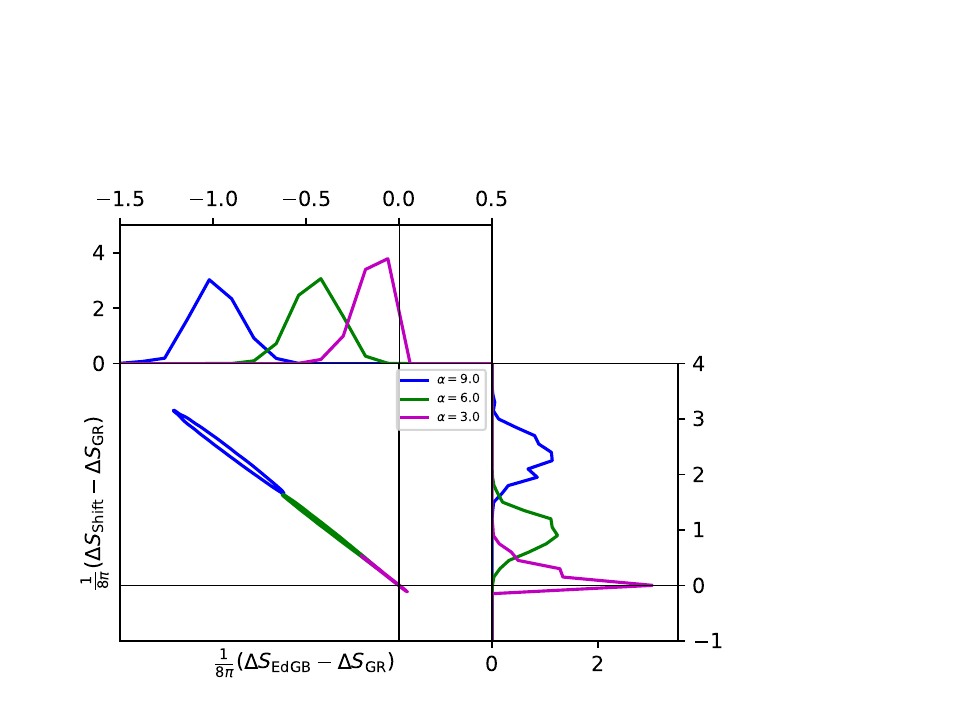}
    \caption{Plot showing the histogram of values of $\delta(\Delta S)$ for the dilaton and the shift symmetric cases, for three different $\alpha$ values, with the BBH parameters fixed to the Case B event. $\alpha$ is expressed in $\mathrm{[km]}^2$ units. The contours are seen to touch the GR baseline for $ \sqrt{\alpha} = \sqrt{3}$ km.}
    \label{fig:cshange}
\end{figure}

In contrast to our results for the simulated cases, we see that the $\delta(\Delta S)$ histograms for the real events are almost completely consistent with GR, with only GW170104 showing a very small deviation which is statistically significant. This means that an injection of $\alpha = 9.0 \mathrm{[km]}^2$ is large enough to be inconsistent with the result one gets from computing $\delta(\Delta S)$ for real events. Consequently at this point, it is noteworthy for us to pause and ask the question that given an event what would be the smallest $\alpha$ which could add a meaningful contribution to the value of $\delta(\Delta S)$. We answer this by employing a second procedure wherein we consider only mock events. We fix the non-$\alpha$ parameters of a set of three injections to Case B, but vary the value of injected $\alpha$, as shown in Fig \ref{fig:cshange}. As per expectation, we clearly see that as $\alpha \rightarrow 0$ the corresponding $\delta(\Delta S)$ contours shift towards the GR baseline $(0,0)$ point and touch it when $ \alpha = \sqrt{3}$ km at which, the results of our simulated event become consistent with those obtained from the real data.

A consequence of this shift in entropy change is that the presence of $\alpha$ in principle should also change the bounds on the topological coupling parameter $\gamma$. Following the arguments of Ref.~\cite{Chakravarti:2022zeq}, we now see that the global second law now demands
\begin{equation}\label{eq:2ndlaw}
    \delta(\Delta S_\iota) + (\Delta S_\mathrm{GR}) - 4\pi\gamma \geq 0
\end{equation}
where $\iota$ stands for either `EdGB' or `Shift'. From Eq.~\eqref{eq:2ndlaw} it is clear that the $\delta(\Delta S)$ term should add a small perturbative correction to the bound on $\gamma$ which was previously obtained from just the Bekenstein-Hawking entropy change.
However, the exercises performed above highlight a couple of important points. We see that considerations of real data mean that $\alpha \lesssim \sqrt{3}$ [km]. We also see that at such $\alpha$ values $\delta(\Delta S_\iota) \simeq 0$ implying that events involving $\mathcal{O}(10) M_\odot$ BHs cannot modify the bounds on $\gamma$ at $90\%$ confidence over the baseline GR when $\sqrt{\alpha} \leq \sqrt{3} $[km]. Thus, considering the latest obtained bounds on $\sqrt{\alpha} \lesssim 1.2 $[km] as found in \cite{Lyu:2022gdr}, it is unlikely that the bound on $\gamma$ will improve significantly over the GR case. We remark that observation of binary mergers $\mathcal{O}(1) M_\odot$ BHs may produce observable effects even with the current constraints with $\alpha$, and as such have a better chance of modifying the $\gamma$ bound.

\begin{figure}[ht!]
    \centering
    \includegraphics[width=1.2\columnwidth]{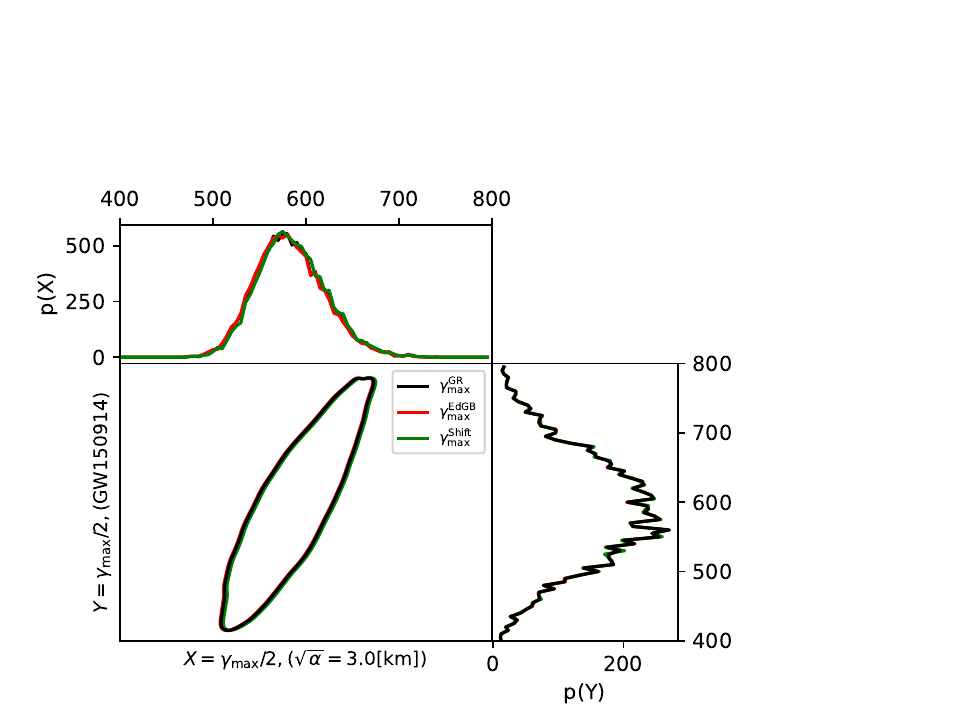}
    \caption{Plot showing the joint histogram of values of $\gamma_\mathrm{max}/2$ for the Case B event with injected $\sqrt{\alpha} = 3.0$[km] (x-axis) and the realistic GW150914 event (y-axis). See text for details.}
    \label{fig:gammax}
\end{figure}

Before concluding this section, it is worthwhile to explicitly see how much an unrealistic value of $\alpha$ pushes $\gamma_\mathrm{max}$ following Eq.~\eqref{eq:2ndlaw}, and compare it to a realistic scenario. Accordingly, we jointly show the histograms of $\gamma_{\mathrm{max}}$ from Case B and GW150914 in Fig.~\ref{fig:gammax}. We observe that even the unphysical $\sqrt{\alpha} = 3.0$[km] only has a very small shift to $\gamma_\mathrm{max}$ which is nevertheless noticeable. As expected we also see $\alpha$ to have no effect at all on the value of $\gamma_{max}$ for the GW150914 event.

\subsection{Measure of bias and its dependence on signal SNR}\label{ssec:bias} 

\begin{figure}[ht!]
    \centering
    \includegraphics[width=1.2\columnwidth]{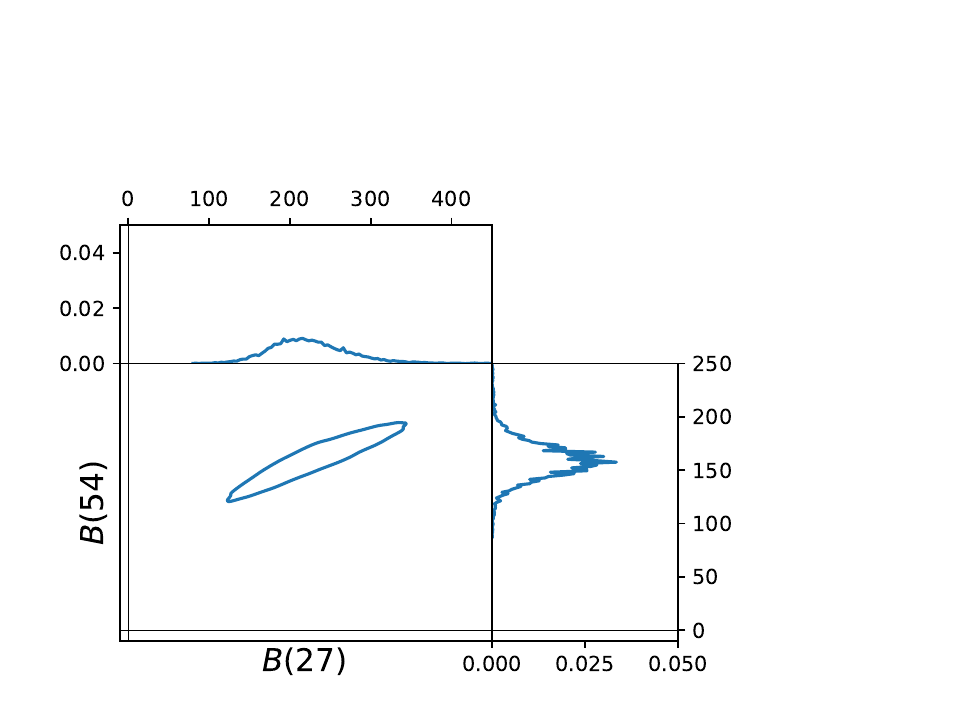}
    \caption{Plot showing the histogram of values of $B(\rho)$ for the events tabulated in Table \ref{Table:revents}. Note that the measure of bias gets sharper with the higher SNR of $\rho = 54$   }
    \label{fig:rcnge}
\end{figure}

The results of Sec.~\ref{ssec:alpha}, demonstrate that the effect of $\alpha$ on $\delta(\Delta S)$ and therefore $\gamma_{\mathrm{max}}$ is vanishingly small and one needs to inject unphysically large $\alpha$ to get any meaningful deviation over the GR baseline. However it should also be noted that the GW170104 event shows a very small deviation. In this section, we aim to investigate if template biases could be responsible for the deviation thus observed. For this we just consider the bias that is produced at the level of the BH entropy difference. In other words, we want to estimate $\Delta S_\mathrm{GR}$ for a pair of merging BHs at different values of SNR $\rho$, and then compare them against the true value $\Delta S^0_\mathrm{GR}$ which would then give the measure of bias
\begin{equation}\label{eq:bias}
    B(\rho) = \frac{1}{8\pi}(\Delta S^\rho_\mathrm{GR} - \Delta S^0_\mathrm{GR})
\end{equation}
as a function of SNR. Accordingly we have computed the inference of necessary parameters using this time a GW200225-like event \cite{abbott2023gwtc} for two different instances of SNR. The details of the parameters are shown in Table~\ref{Table:revents}. Accordingly, we have simulated a set of two GW200225-like events with the `LIGO-HL' network SNR of $21$ and $54$ respectively. The resulting biases as defined in Eq.~\eqref{eq:bias} have been plotted in Fig.~\ref{fig:rcnge}.

\begin{table}
\begin{center}
\begin{tabular}{||c| c| c| c| c| c||} 
 \hline
 \# &  $m_1$ & $m_2$ & $\chi_1$ & $\chi_2$ & $\rho$(SNR) \\ [0.5ex] 
 \hline\hline
 1& 19.3 & 14.0 & -0.14 & -0.08 & 27 \\ 
 \hline
 2& 19.3 & 14.0 & -0.14 & -0.08 & 54 \\
 \hline
\end{tabular}
\end{center}
\caption{Table showing the configurations for our choices of a GW200225-like event, injected at two SNR values. As before the SNRs are LIGO-HL network SNRs.}
\label{Table:revents}
\end{table} 

Unsurprisingly we see that with increasing SNR the histograms become narrower, which allows us to better visualise the magnitude of the absolute bias. This absolute bias is again seen to be of $\sim \mathcal{O}(100)$. Working out the numbers we see that this is $\sim 30-40\%$ of $\Delta S_\mathrm{GR}$ itself. To explain this, we recall that BBH GW templates cannot measure component masses very accurately and have been previously found to introduce systematic errors in component masses in the $\sim 15-20\%$ ballpark \cite{Chakravarti:2018uyi}. As $\Delta S_\mathrm{GR}\sim M^2$, we see that the bias in component mass alone is able to account for almost all of the total bias. We should also remember that spin estimation would add another source of bias to the error budget.  Returning to the problem at hand we see that template biasing is more than sufficient to explain the deviation seen with the GW170104 event. Following  Eq.~\eqref{eq:2ndlaw}, we can clearly see that this same $30-40\%$ bias  will also affect the bound on $\gamma$ and is clearly at least an order of magnitude greater than any effect produced by $\alpha$.  
\section{Summary \& Conclusions}

The presence of a topological coupling $\kappa \gamma \, \mathcal{G}$ in the gravitational action can influence black-hole thermodynamics through its contribution to their entropy, impacting for example the BH nucleation rate in de Sitter which can lead to instabilities. A bound on this coupling can be obtained from demanding the validity of the global second law of thermodynamics during binary black-hole mergers. Building on previous literature, we aimed to study those factors that could push the bounds on the topological coupling $\gamma$ away from its GR value when the BHs carry scalar hair. To do that we made use of metric solutions and waveforms in a perturbative framework to compute the Wald entropy, a natural generalisation of the Bekenstein-Hawking result for beyond Einstein-Hilbert actions, for two instances of a scalar-Gauss-Bonnet coupling, namely the EdGB and linear-sGB theories. The perturbative nature of the coupling $\alpha$ in relation to the BH mass $M$ ensures the corrections to be small. Even though we show that a statistically significant deviation in the entropy at values of $\alpha \geq 9\mathrm{[km]}^2$ is possible, no such deviation shows up in the real data from our list of selected GW events. This leads us to conclude that with observation of $\mathcal{O}(10)$ $M_\odot$ BHs, the $\alpha$ effects on entropy become statistically indistinguishable from GR at $\alpha \simeq 3.0 \mathrm{[km]}^2$, a value which is already ruled out by the current bounds on $\alpha\lesssim (1.2)^2 \mathrm{[km]}^2$ obtained from inspiral dynamics. Going to smaller BH masses of a few $M_\odot$ might offer us a way out, since ultimately effects depend on the ratio $\alpha/M^2$. Caution must be taken, however, as this also brings the BHs closer to the cutoff of the EFT, $\Lambda_{UV} < 1/\sqrt{\alpha}$, signaling new-physics effects becoming important \cite{Serra:2022pzl}. Moreover, this would also require going beyond the perturbative regime which necessitates full numerical relativity simulations of these extended theories, which will be an interesting aspect to explore in the future. In this context, we also find that GW template biases shown in Fig.~\ref{fig:rcnge} have effects that are two orders of magnitude greater than those of the coupling $\alpha$, and thus have a much bigger risk of pushing $\gamma$ away from its true value. It can be safely concluded that biases from GW templates should be focused on with priority during any future work in this direction.

It is also worthwhile to note that in principle, our computation and subsequent estimation of entropy can distinguish between a dilaton-type or a shift-symmetric action. However, this distinction is not a possibility at current state-of-art simply because entropy cannot be directly observed. We believe that the direct measurement of BH temperature would break this deadlock, from which our results could be meaningfully used to test the nature of the scalar coupling.

\section*{Acknowledgments}
The authors acknowledge Kent Yagi for providing necessary Mathematica scripts. The authors also thank David Kubizňák, Sudipta Sarkar and Rajes Ghosh for helpful discussions and Josef Dvořáček for necessary IT support on the CEICO cluster. K.C acknowledges research support by the PPLZ grant (Project number:10005320/ 0501) of the Czech Academy. The work of L.G.T.\ was supported by the European Union (Grant No.\ 101063210). A.R is supported by the research program of the Netherlands Organisation for Scientific Research (NWO). A.R gratefully acknowledges the Central European Institute of Cosmology (CEICO), Prague, Czech Republic, for their generous hospitality during his visit.

\appendix

\section{Entropy for a Kerr-like hairy black hole}\label{app:entropy}

In order to compute the entropy for EdGB theory from Wald's entropy formula given by Eq.~\eqref{Wald-entropy-EdGB}, we rely on the analytic black-hole solution computed perturbatively around Kerr in Ref.~\cite{Ayzenberg:2014aka}. This solution is accurate up to $\mathcal{O}(\alpha,\chi^2)$ for $\phi$, and $\mathcal{O}(\alpha^2,\chi^2)$ for the metric $g_{\mu\nu}$. At the level of the entropy formula Eq.~\eqref{Wald-entropy-EdGB}, the precision is set by the Bekenstein-Hawking term $A_\mathcal{H}/4$, which is computed from the metric solution $g_{\mu\nu}$ and thererfore it is also accurate up to $\mathcal{O}(\alpha^2,\chi^2)$, as seen in Eq.~\eqref{eq:horizon-area}. For this reason, for perturbative consistency we only need to compute the new contribution arising from the EdGB term up to the same precision, i.e. $\mathcal{O}(\alpha^2,\chi^2)$. Moreover, this contribution in itself already contains a factor of $\alpha$ and a factor of $\phi$,
\begin{eqnarray} \label{entropy-alpha}
    I \equiv \frac{\alpha}{4\kappa} \int_{\mathcal{H}} \phi \left( \epsilon^{\mu\nu} R - 4 \epsilon^{\lambda \mu} R_\lambda{}^{\nu} + \epsilon_{\lambda\rho} R^{\nu\mu\lambda\rho} \right) \epsilon_{\mu\nu} d^2 \! A \,, \qquad
\end{eqnarray}
making it only necessary to compute the rest of the integrand at $\mathcal{O}(\alpha^0,\chi^2)$ to achieve said precision. This means in practice utilizing the unperturbed Kerr expressions for the Riemann tensor $R^{\mu}{}_{\nu\rho\sigma}$ and its contractions, the binormal tensor $\epsilon_{\mu\nu}$ and the area element $d^2\! A =  \sqrt{g_{\theta\theta} g_{\varphi\varphi}}|_{\mathcal{H}} \, d\theta d\varphi$, where $\theta$ and $\varphi$ are the standard angular variables on the 2-sphere. An immediate consequence of this is that, since Kerr is a vacuum solution of GR, it holds that $R_{\mu\nu} = 0$ (and thus also $R = 0$), leaving only a single term to be computed.

Since we want to compute the result up to $\mathcal{O}(\alpha^2)$, and given that $\phi = \mathcal{O}(\alpha)$, it is sufficient to evaluate the expression between brackets at $\mathcal{O}(\alpha^0)$, which is simply the unperturbed Kerr metric. In this case, since Kerr is a vacuum solution of GR, we have $R_{\mu\nu} = 0$, and therefore only the last term (Riemann) is nonvanishing,
\begin{eqnarray}
    I &=& \frac{\alpha}{4\kappa} \int_{\mathcal{H}} d\theta d\varphi \sqrt{g_{\theta\theta} g_{\varphi\varphi}} \phi \, \epsilon_{\mu\nu} \epsilon_{\lambda\rho} R^{\nu\mu\lambda\rho} \Big|_{\mathcal{H},\mathrm{Kerr}} + \mathcal{O}(\alpha^3) \, , \notag \\
\end{eqnarray}
where, as mentioned, it suffices to use the Kerr solution at this order.

The Kerr metric is given in the Boyer-Lindquist coordinates (in the notation of \cite{Chandrasekhar:1985kt}) as
\begin{eqnarray}
    ds^2 &=& - \rho^2 \frac{\Delta}{\Sigma^2} dt^2 + \frac{\Sigma^2}{\rho^2} \left( d\varphi - \frac{2aMr}{\Sigma^2} dt \right)^2 \sin^2(\theta) \notag \\
    &&+ \frac{\rho^2}{\Delta} dr^2 + \rho^2 d\theta^2,
\end{eqnarray}
where
\begin{eqnarray}
    \Delta &=& r^2 - 2 M r + a^2, \\
    \rho^2 &=& r^2 + a^2 \cos^2(\theta), \\
    \Sigma^2 &=& \rho^2 (r^2 + a^2) + 2 a^2 M r \, \sin^2(\theta),
\end{eqnarray}
and $a$ is the spin parameter. The event horizon corresponds to the largest root of $\Delta = 0$, i.e. $r_H = M + (M^2 - a^2)^{1/2}$, and has an area of
\begin{eqnarray}
    A_{\mathcal{H},\mathrm{Kerr}} = 4\pi (r_H^2 + a^2) \, .
\end{eqnarray}
This metric has a Killing vector field $\xi^\mu = t^\mu + \Omega_\mathcal{H} \, \varphi^\mu$, where $\Omega_\mathcal{H}$ is the \emph{constant} angular velocity of the horizon. From this, we can define the binormal tensor as
\begin{eqnarray}
    \epsilon_{\mu\nu} = \kappa_\mathcal{H} \nabla_{[\mu} \xi_{\nu]} = \kappa_\mathcal{H} \left( \partial_\mu \xi_\nu - \partial_\nu \xi_\mu \right) \, ,
\end{eqnarray}
 where $\kappa_\mathcal{H}$ is the horizon surface gravity, which normalizes it to $\epsilon_{\mu\nu} \epsilon^{\mu\nu} = -2$.

A long but straight-forward computation leads to the desired result at the required precision,
\begin{eqnarray}
    && \sqrt{g_{\theta\theta} g_{\varphi\varphi}} \, \epsilon_{\mu\nu} \epsilon_{\lambda\rho} R^{\mu\nu\lambda\rho} \Big|_{\mathcal{H},\mathrm{Kerr}} \notag \\
    &=& \sin(\theta) \left[ 1 + \frac{\chi^2}{2} \left(1 - 3 \cos(\theta)^2\right) \right] + \mathcal{O}(\chi^4) \, .
\end{eqnarray}
where we have also expanded for small $\chi = a/M$, i.e. the dimensionless spin parameter. The above expression needs to be combined with the scalar field solution from \cite{Ayzenberg:2014aka}, valid up to $\mathcal{O}(\chi^2,\alpha)$, evaluated at the horizon
\begin{eqnarray}
    \phi_\mathcal{H} = \frac{\alpha}{M^2} \left[\frac{11}{6} - \chi^2 \frac{{\left(118 \, \cos(\theta)^2 - 25\right)}}{80} \right] \, .
\end{eqnarray}

We can now compute the explicit $\alpha$ contribution to the entropy, Eq.~\eqref{entropy-alpha} up to order $\mathcal{O}(\chi^2,\alpha^2)$,
\begin{eqnarray}
    I &\simeq& - \frac{2 \pi \alpha}{\kappa} \int_{0}^\pi d\theta \sqrt{g_{\theta\theta} g_{\varphi\varphi}} \, \phi_\mathcal{H}  \epsilon_{\mu\nu} \epsilon_{\lambda\rho} R^{\mu\nu\lambda\rho} \Big|_{\mathcal{H},\mathrm{Kerr}} \notag \\
    &=& \frac{4 \pi \alpha^2}{\kappa M^2} \left( \frac{11}{6} - \chi^2 \frac{43}{240} \right) \, .    
\end{eqnarray}
which gives the result of Eq.~\eqref{eq:Dltn_final} for the EdGB entropy. Notice the index order swap with respect to Eq.~\eqref{entropy-alpha}, which accounts for the minus sign. 

Both entropy formulas for the EdGB and linear-sGB cases, Eqs.~\eqref{eq:Dltn_final} and \eqref{eq:Shft_final} respectively, can be combined into
\begin{eqnarray}
    S_\mathcal{H} &=& 4\pi M^2 \left( 1 - \frac{\chi^2}{4} + \frac{\alpha^2}{k M^4} \beta(\chi) \right) + 4\pi \gamma \, ,
\end{eqnarray}
where we have made the whole of the $\alpha$-dependence fully explicit, including the one inside $A_\mathcal{H}$ (see Eq.~\eqref{eq:horizon-area}) and we are also defining 
\begin{equation}
    \beta(\chi) = 
    \begin{cases} 
    \frac{73}{120} - \frac{53}{480} \chi^2 \, , \quad &\text{EdGB} ,\\
    -\frac{49}{40} + \frac{11}{160} \chi^2 \, , \quad &\text{linear-sGB} .
    \end{cases}
\end{equation}

\bibliographystyle{unsrt}
\bibliography{refs}

\end{document}